\documentclass[reprint,
superscriptaddress,
amsmath,amssymb,
apl,
]{revtex4-2}

\usepackage[usenames,dvipsnames]{color}
\usepackage{graphicx}
\graphicspath{{./images/} }
\usepackage{textcomp}
\usepackage{dcolumn}
\usepackage{bm}
\usepackage{comment}
\usepackage{upgreek}
\usepackage[usenames,dvipsnames]{color}
\usepackage{url}
\usepackage[
bookmarks=true,
pdftex,
hidelinks,
bookmarks=true,
linktocpage=true, 
hyperindex=true
]{hyperref}

\usepackage[all]{hypcap}

\usepackage[mathlines]{lineno}
\usepackage{multirow}
\usepackage{siunitx}
\usepackage{lipsum}
\usepackage{amsmath}

\usepackage{comment}

\DeclareGraphicsExtensions{.pdf,.png,.jpg}

\usepackage{tabularx}
\newcolumntype{Y}{>{\centering\arraybackslash}X}


\usepackage{float}
\usepackage{placeins}

\begin{document}
\title{Minimizing the discrimination time for quantum states of an artificial atom}

\author{I. Takmakov}
\email{ivan.takmakov@kit.edu}
\affiliation{Physikalisches Institut, Karlsruhe Institute of Technology, 76131 Karlsruhe, Germany}
\affiliation{IQMT, Karlsruhe Institute of Technology, 76344 Eggenstein-Leopoldshafen, Germany}
\affiliation{Institute of Nanotechnology, Karlsruhe Institute of Technology, 76344 Eggenstein Leopoldshafen, Germany}
\affiliation{Russian Quantum Center, National University of Science and Technology MISIS, 119049 Moscow, Russia}

\author{P. Winkel}%
\thanks{I. Takmakov and P. Winkel contributed equally to this work.}
\affiliation{Physikalisches Institut, Karlsruhe Institute of Technology, 76131 Karlsruhe, Germany}
\affiliation{IQMT, Karlsruhe Institute of Technology, 76344 Eggenstein-Leopoldshafen, Germany}

\author{F. Foroughi}
\affiliation{CNRS, Institut NEEL, F-3800 Grenoble, France}
\affiliation{Universite Grenoble Alpes, Institut NEEL, F-3800 Grenoble, France}

\author{L. Planat}
\affiliation{CNRS, Institut NEEL, F-3800 Grenoble, France}
\affiliation{Universite Grenoble Alpes, Institut NEEL, F-3800 Grenoble, France}

\author{D. Gusenkova}
\affiliation{Physikalisches Institut, Karlsruhe Institute of Technology, 76131 Karlsruhe, Germany}

\author{M. Spiecker}
\affiliation{Physikalisches Institut, Karlsruhe Institute of Technology, 76131 Karlsruhe, Germany}

\author{D. Rieger}
\affiliation{Physikalisches Institut, Karlsruhe Institute of Technology, 76131 Karlsruhe, Germany}

\author{L. Grünhaupt}
\affiliation{Physikalisches Institut, Karlsruhe Institute of Technology, 76131 Karlsruhe, Germany}

\author{A. V. Ustinov}
\affiliation{Physikalisches Institut, Karlsruhe Institute of Technology, 76131 Karlsruhe, Germany}
\affiliation{Russian Quantum Center, National University of Science and Technology MISIS, 119049 Moscow, Russia}

\author{W. Wernsdorfer}
\affiliation{Physikalisches Institut, Karlsruhe Institute of Technology, 76131 Karlsruhe, Germany}
\affiliation{IQMT, Karlsruhe Institute of Technology, 76344 Eggenstein-Leopoldshafen, Germany}
\affiliation{Institute of Nanotechnology, Karlsruhe Institute of Technology, 76344 Eggenstein Leopoldshafen, Germany}

\author{I. M. Pop}
\affiliation{Physikalisches Institut, Karlsruhe Institute of Technology, 76131 Karlsruhe, Germany}
\affiliation{IQMT, Karlsruhe Institute of Technology, 76344 Eggenstein-Leopoldshafen, Germany}
\affiliation{Institute of Nanotechnology, Karlsruhe Institute of Technology, 76344 Eggenstein Leopoldshafen, Germany}

\author{N. Roch}
\affiliation{CNRS, Institut NEEL, F-3800 Grenoble, France}
\affiliation{Universite Grenoble Alpes, Institut NEEL, F-3800 Grenoble, France}

\date{\today}

\begin{abstract}
Fast discrimination between quantum states of superconducting artificial atoms is an important ingredient for quantum information processing. 
In circuit quantum electrodynamics, increasing the signal field amplitude in the readout resonator, dispersively coupled to the artificial atom, improves the signal-to-noise ratio and increases the measurement strength.
Here we employ this effect over two orders of magnitude in readout power, made possible by the unique combination of a dimer Josephson junction array amplifier with a large dynamic range, and the fact that the readout of our granular aluminum fluxonium artificial atom remained quantum-non-demolition (QND) at relatively large photon numbers in the readout resonator, up to  $\overline{n} = 110$.
Using Bayesian inference, this allows us to detect quantum jumps  faster than the readout resonator response time $2/\kappa$, where $\kappa$ is the bandwidth of the readout resonator.
\end{abstract}
\maketitle
Quantum jumps --- transitions between discrete energy levels of a quantum system --- have been measured over the last two  decades in various physical systems \cite{Nagourney1986,Neumann2010,Robledo2011,Gleyzes2007,Jelezko2002,Vijay2011}.
The measurement of quantum jumps relies on the ability to discriminate the individual states of a quantum system on a timescale significantly shorter than their energy relaxation times.
For applications in quantum technologies, this ability is instrumental in the implementation of quantum error correction algorithms \cite{Reed2012,Riste2013a,Sun2014,Kelly2015,Ofek2016}, a milestone on the quantum information processing roadmap, or can be used as a detection tool informing on the interactions between the quantum system and its environment \cite{Riste2013,Vool2014,Serniak2019}.
For all these applications it is important to discriminate between the states of a quantum system as fast as possible.

The speed of quantum-non-demolition (QND) readout in circuit quantum electrodynamics \cite{Wallraff2004,Krantz2019a,Blais2020a,Kjaergaard2020} improved significantly over the last decade \cite{Jeffrey2014,Walter2017,Dassonneville2020}, mainly thanks to the development of near-quantum limited Josephson parametric amplifiers \cite{Castellanos-Beltran2007,Yamamoto2008,Mutus2014,Roch2012,Eichler2014,Macklin2015,Roy2016}, which increase the signal-to-noise ratio (SNR) by reducing the noise of the measurement setup down to the quantum limit \cite{Caves1982}.
An additional resource to speed-up the measurement, the increase of the readout drive power has yet barely been exploited, because superconducting qubits are observed to suffer from increased energy relaxation and leakage out of computational space when the circulating photon number in the resonator, $\overline{n}$, is increased.
In practice, these effects bound the photon number for QND readout \cite{Johnson2012,Sank2016,Walter2017,Minev2019}, typically to $\overline{n} < 10$, for reasons which are currently the subject of theoretical investigation \cite{Boissonneault2009,Malekakhlagh2020,Petrescu2020}.

By performing a continuous measurement and benefiting from the remarkable insensitivity to $\overline{n}$ of a granular Aluminum (grAl) fluxonium \cite{Grunhaupt2019} artificial atom, recently reported in Ref.\cite{Gusenkova2020}, we demonstrate a significant decrease of the QND state discrimination time with increasing $\overline{n}$, up to $\overline{n} = 110$.
To handle the correspondingly large signal power we use a high dynamic range dimer Josephson junction array parametric amplifier (DJJAA) \cite{Winkel2019}.
We are able to reduce the mean time required to discriminate between the qubit states from 1.2~\textmu s at $\overline{n} = 1.7$ photons, down to 0.175~\textmu s at $\overline{n} = 56$ photons, which is shorter than the time required for the resonator pointer state to transition between the steady states corresponding to the qubit population.
This discrimination time is not limited by the integration time for the quadrature decomposition of the readout signal, which, in principle can be made arbitrarily short, but by the onset of squeezing caused by the nonlinearity of the readout resonator, and the increasing excitation of higher energy eigenstates of the artificial atom. 
Since the integration time is significantly shorter than the $\tau_B = 290$~ns response time of the readout resonator, we employ Bayesian filtering \cite{Sun2014,Korotkov2016,Wang2015,Weber2016a,Feng2016,Reick2010,Six2016} to detect quantum jumps before the system reaches a new steady state.

\begin{figure*}
	\includegraphics[width=1 \linewidth] {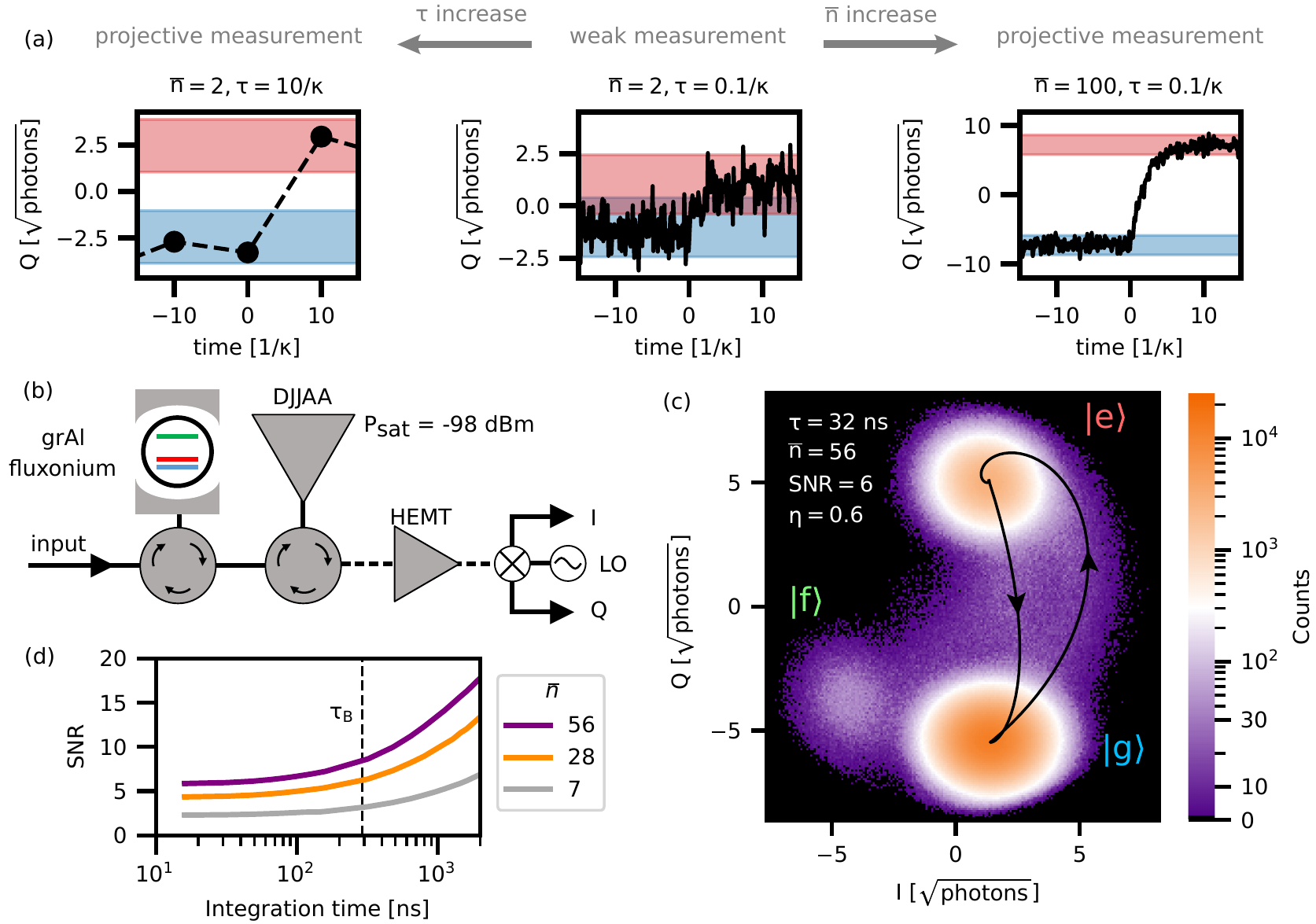}
	\caption{ 
	    (a) Simulated Q quadrature (black lines) of a quantum-limited continuous measurement of a qubit coupled to a readout resonator.
		The quadrature values are presented in  units of square root of measurement photons $\sqrt{\bar{n} B^{-1}\kappa/4}$, where $\overline{n}$ is the circulating photon number in the readout resonator, $\kappa$ is the resonator's coupling rate to the measurement apparatus, and $B^{-1} \approx \tau + \tau_B$ is the measurement bandwidth, given by the sum of the integration time $\tau$ and the resonator response time $\tau_B = 2/\kappa$ (see appendix \ref{section:DJJAA}).
		The blue and red areas indicate $\pm 2 \sigma$ intervals centered around the mean values of the Q quadrature corresponding to the ground and first excited state, respectively, denoted $Q_{|\text{g}\rangle}$ and $Q_{|\text{e}\rangle}$.
		The panels depict the transition from a weak measurement (center) to a projective measurement by increasing either $\tau$ (left panel) or $\overline{n}$ (right panel).
		(b) Simplified schematic of the experimental setup. 
		The incident readout signal reflects from a resonator coupled to a fluxonium artificial atom, both implemented using granular aluminum, which is the same device discussed in Ref.\cite{Gusenkova2020}. 
	    The reflected signal is amplified by a dimer Josephson junction array amplifier (DJJAA) \cite{Winkel2019} operated in the non-degenerate phase-preserving regime.
		The measured 1~dB compression point of the DJJAA at 20~dB of power gain is $P_{\text{sat}} = -98$~dBm (see appendix \ref{section:DJJAA}).
		The signal is further amplified by a commercial high electron mobility transistor amplifier (HEMT) thermalized at 4~K, and is demodulated at room temperature into the I and Q quadratures.
		(c) 2D histogram of continuously measured I and Q quadratures, presented in  units of square root of measurement photons $\sqrt{\bar{n} B^{-1}\kappa/4}$, where $\bar{n} = 56$, and $B^{-1} = 322$~ns. 
		In order to highlight the resonator classical trajectories, the colorbar scale is linear below 30 counts and logarithmic above. 
		The calculated trajectories ($|\text{g}\rangle$ to $|\text{e}\rangle$ and back) corresponding to the linear response of the resonator are indicated by the black traces.
		The signal-to-noise ratio, defined by Eq.\eqref{eq:SNR}, is SNR = 6, and the obtained quantum efficiency for individual pointer state measurements is $\eta = 0.6 \pm 0.1$ (see appendix \ref{section:DJJAA}).
		(d) Signal-to-noise ratio measured as a function of integration time $\tau$ for different  $\overline{n}$ values.
		The black dotted line corresponds to $\tau_{\text{B}}$.
}
		\label{fig:fig1}
\end{figure*}

During the continuous measurement of a qubit coupled to a readout resonator, schematically shown in Fig.\ref{fig:fig1}(a), when the qubit state changes between ground ($|\text{g}\rangle$) and excited ($|\text{e}\rangle$), the readout signal quadrature evolves as $Q(t) - Q_{|\text{e}\rangle} \propto \left( Q_{|\text{g}\rangle} - Q_{|\text{e}\rangle}\right)e^{-\kappa t /2}$.
Here, $Q_{|\text{g}\rangle/|\text{e}\rangle}$ are the measured readout resonator quadrature values (see Fig.\ref{fig:fig1}(b)) of the steady states corresponding to $|\text{g}\rangle$ and $|\text{e}\rangle$ (see Appendix \ref{section:DJJAA}).
The resonator bandwidth $\kappa$ sets the response time $\tau_B = 2/\kappa$ of the measurement apparatus to a quantum jump. 
The readout power, populating the readout resonator with $\overline{n}$ photons, and the integration time $\tau$ determine whether a single measured quadrature point is sufficient (projective measurement) or insufficient (weak measurement) to determine the qubit state. 
A typical example of a sequence of weak measurements is shown in the middle panel of Fig.\ref{fig:fig1}(a). 
It can be converted to a strong measurement either by increasing $\tau$ (left panel of Fig.\ref{fig:fig1}(a)), which unfortunately reduces the time resolution of the qubit state determination, or by increasing $\overline{n}$ (right panel of Fig.\ref{fig:fig1}(a)).
While the quadrature response time of the readout resonator is limited to $\tau_B$, given sufficiently large SNR, the qubit state can be inferred before the quadrature reaches its steady state. 
Therefore, as we will show in the following, increasing $\overline{n}$ is a viable strategy to speed-up qubit state detection, provided the qubit and the measurement chain can handle the increased readout power.

Our experimental setup is shown in Fig.\ref{fig:fig1}(b).
The monitored quantum system is a fluxonium artificial atom \cite{Manucharyan2009}, dispersively coupled via a shared inductance to a readout resonator (see Table~\ref{tab:table1} for the list of system parameters).
The inductors required for both the fluxonium and readout antenna are implemented using grAl \cite{Grunhaupt2019}, and the fluxonium junction is implemented in the shape of a SQUID \cite{Gusenkova2020,Lin2018}.
The sample chip was placed in a rectangular waveguide, following the concept in Ref.\cite{Kou2018}, and measured in reflection (see Fig.\ref{fig:fig1}(b)).
We used a DJJAA \cite{Winkel2019} to perform  non-degenerate,  phase-preserving amplification \cite{Roy2016,Eichler2014a} with 20~dB of power gain and 7~MHz instantaneous bandwidth. 
The DJJAA saturation power was -98~dBm (see appendix \ref{section:DJJAA}), corresponding to a circulating photon number in the readout resonator $\overline{n} \approx 10^4$.

In Fig.\ref{fig:fig1}(c) we show a typical histogram of the demodulated readout resonator response for $\overline{n} = 56$ and an integration time $\tau = 32$~ns.
The distribution shows three maxima, corresponding to the first three lowest energy levels of the fluxonium artificial atom, labeled  $|\text{g}\rangle,|\text{e}\rangle,|\text{f}\rangle$.
Since the integration time $\tau = 32$~ns is much shorter than the characteristic response time of the measurement setup $\tau_B = 290$~ns, trajectories corresponding to transitions between the three steady states are visible, filling the space between the maxima in the IQ-plane.
The steady states are visibly squeezed because of the grAl resonator's intrinsic \cite{Maleeva2018} and inherited nonlinearity (see appendix \ref{section:latching}).
\begin{table}[t]
	\begin{center}
		\caption{
		List of the qubit-resonator circuit parameters.
		The Josephson energy $E_J$  is in-situ flux-tunable via the SQUID junction.
		The flux bias for the fluxonium qubit was fine-tuned in the vicinity of $\Phi_0/2$ to maximize pointer states separation.
		}
		\label{tab:table1}
		\begin{tabularx}{\columnwidth}{lYY}
	\hline
	$E_C/h$, GHz &  2.8 \\ 
    $E_L/h$, GHz &    0.71	\\ 
    $E_{\text{J}}/h$, GHz	&   7.1  \\
    Fluxonium flux bias, $\Phi_0$	&   0.542 \\
	{$\kappa/2\pi$, MHz}	&	1.1	\\ 
	$f_{\text{res}}$, GHz	&	7.247 \\ 
	$\chi_{\text{ge}}/2\pi$, MHz 	&	-1.09 \\ 
	$f_{\text{ge}}$, MHz	&	902	\\ 
	$f_{\text{ef}}$, MHz &	7701	\\ 
	$T_{1}$, \textmu s&	$20 \pm 4$ \\ \hline
		\end{tabularx}
	\end{center}
\end{table}
 \begin{figure}[t]
	\includegraphics[width=1. \linewidth] {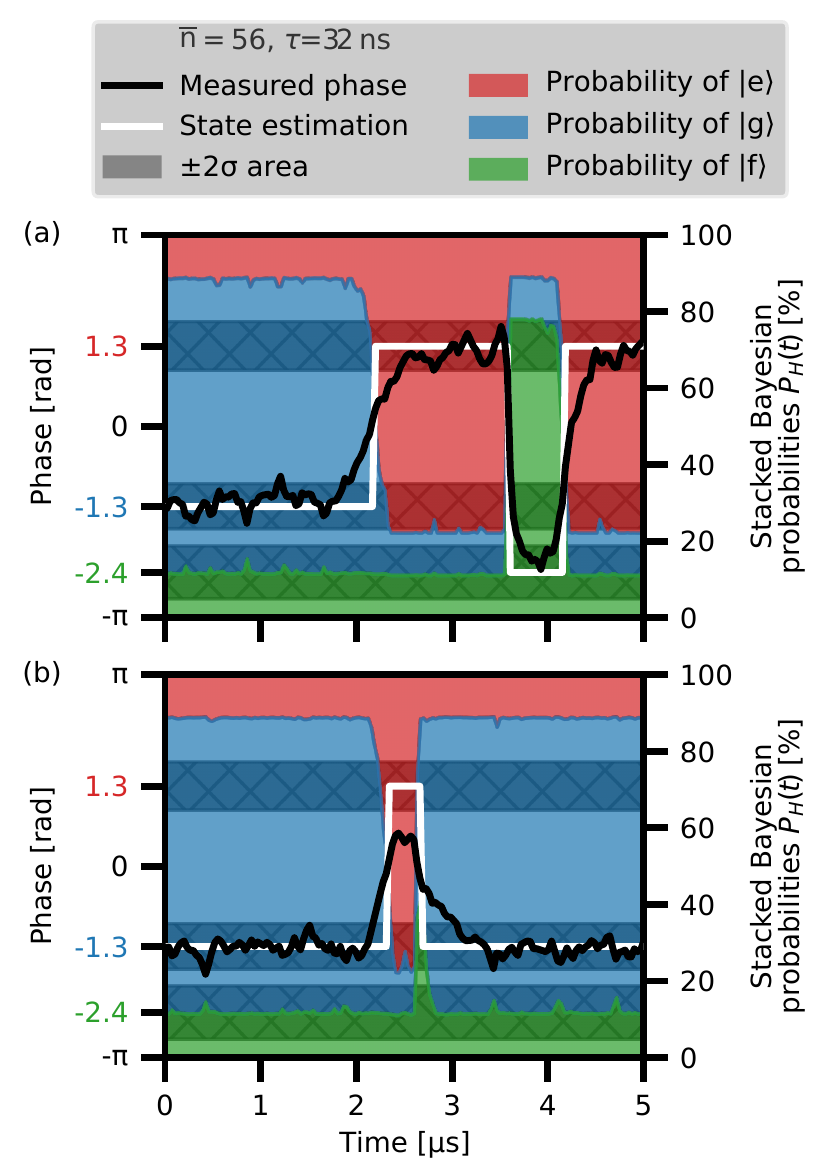}
	\caption{
		Examples of Bayesian inference of the artificial atom's state ( $|\text{g}\rangle$, $|\text{e}\rangle$, or $|\text{f}\rangle$ ) based on the continuous wave measurement of the readout resonator.
		Black lines represent the phase vs. time extracted from the measured IQ response (see Fig.\ref{fig:fig1}b) for $\overline{n} = 56$ and $\tau = 32$~ns.
		Hatched areas indicate the $\pm 2 \sigma$ intervals for each state, centered on the corresponding average phase response: -1.3 for $|\text{g}\rangle$ (blue label),  1.3 for $|\text{e}\rangle$ (red label), and -2.4 for $|\text{f}\rangle$ (green label).
		Colored areas show the stacked barplot vs. time for the calculated Bayesian probabilities (see Eq.\eqref{Bayesian_formula}) of the first three states of the artificial atom: $|\text{g}\rangle$ in blue, $|\text{e}\rangle$ in red, and $|\text{f}\rangle$ in green.
		The inferred  quantum state is indicated by the white line; when one of the Bayesian probabilities reaches 50\%, a jump to the respective state is declared.
	}
	\label{bayesian_result}
\end{figure}
In Fig.\ref{fig:fig1}(d) we plot the measured SNR, defined as the ratio between the separation of the pointer states in the IQ plane and their variance,  
\begin{equation}
\text{SNR}= \frac{\left|\bm{\alpha_{|\text{e}\rangle}} - \bm{\alpha_{|\text{g}\rangle}}\right|}{\sigma_{|\text{e}\rangle} + \sigma_{|\text{g}\rangle}} = \sqrt{\frac{1}{4}\kappa \eta \overline{n}(\tau + \tau_{\text{B}})}\sin{\frac{\varphi_{\text{eg}}}{2}},
\label{eq:SNR}
\end{equation}
 versus the integration time $\tau$, where $\bm{\alpha_{|\text{g}\rangle,|\text{e}\rangle}}$ are the pointer states, $\sigma_{|\text{g}\rangle,|\text{e}\rangle}$ are the corresponding variances, and $\varphi_{eg}$ is the phase difference between the pointer states, $\eta = 0.6 \pm 0.1$ is the quantum effeciency (see appendix \ref{section:DJJAA}).
Following Eq.\eqref{eq:SNR}, the SNR saturates when $\tau$ goes below~$\tau_{\text{B}}$.
The resonator response time can therefore significantly slow down  the detection of quantum jumps based on fixed IQ thresholds, commonly used for latching filters (see Appendix~\ref{section:latching}).
However, since the state of the artificial atom is encoded in the instantaneous evolution of the pointer state, given sufficient SNR it can be extracted even before the resonator reaches its steady state.
Thanks to the increased SNR made available by measurements at large $\overline{n}$, as shown in Fig.\ref{fig:fig1}(d), we can implement a recursive Bayesian filter for quantum jumps detection, assuming a hidden Markov model, similarly to Refs.\cite{Sun2014,Korotkov2016,Wang2015,Weber2016a,Feng2016,Reick2010,Six2016}: 
\begin{figure*}[t]
	\includegraphics[width=1. \linewidth]{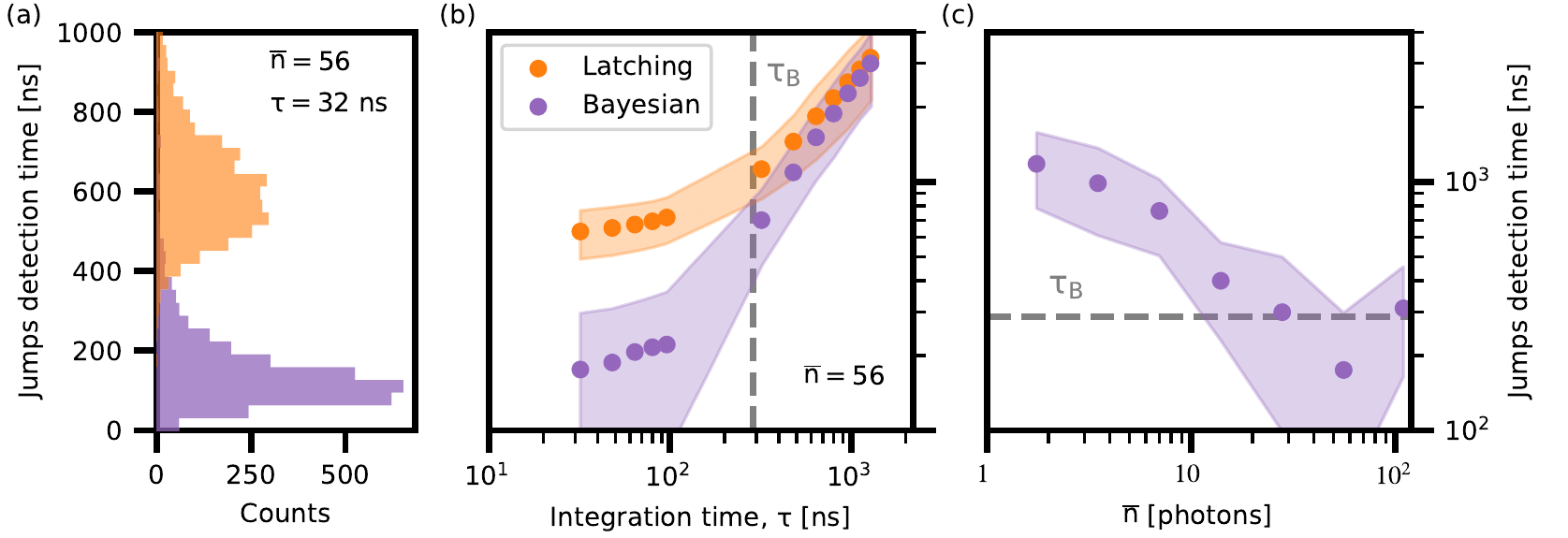}
	\caption{
		(a) Histograms of the quantum jump detection times obtained with a three-point latching filter (orange, see appendix \ref{section:latching}, and a recursive Bayesian filter (purple, see Eq.\ref{Bayesian_formula}), for $|\text{g}\rangle \rightarrow |\text{e}\rangle$.
		For this histogram $\overline{n} = 56$, and the integration time is $\tau = 32$~ns.
		(b) Average detection time for $|\text{g}\rangle \rightarrow |\text{e}\rangle$ jumps as a function of integration time, obtained using the three-point latching filter (in orange) or the recursive Bayesian filter (in purple).
		Colored areas represent the standard deviation of the jump detection time distribution (see panel a).
		(c) Quantum jumps detection time as a function of photon number in the readout resonator $\overline{n}$, for $|\text{g}\rangle \rightarrow |\text{e}\rangle$ quantum jumps.
	    The colored area represents the standard deviation of the jump detection time distribution (see panel a).
		}
	\label{fig:detection_time_histos}
\end{figure*}
\begin{multline}
P_H\left( t+ \tau \right)
= \frac{P\left( \varphi_{t + \tau},\varphi_{t}| {H}\right) \times P_H\left(t\right)}{\sum\limits_{H' \in \{\text{g,e,f}\}} P\left( \varphi_{t + \tau},\varphi_{t}|H' \right)\times P_{H'}(t)  }
\label{Bayesian_formula}
\end{multline}
\begin{multline}
P\left( \varphi_{t + \tau},\varphi_{t}| H\right) = \exp \left[-\frac{\left( \varphi_{t + \tau} - \varphi_{t + \tau}^{\text{calc}}\left(\varphi_t,H\right)\right)^2}{2\beta_H^2\sigma^2_{H}}\right]
\label{Bayesian_formula_explanation}
\end{multline}
Here, $P_H\left( t+ \tau \right)$ is the conditional probability for each fluxonium state hypothesis $H \in \{\text{g,e,f}\}$, given that pointer state phases $\varphi_{t + \tau}$ and $\varphi_{t}$ were measured at times $t + \tau$ and $t$, respectively.
Equation \eqref{Bayesian_formula} updates the probabilities for each fluxonium state $H$ in time increments of the integration time $\tau$.
Following Eq.\eqref{Bayesian_formula_explanation}, the unnormalized probability $P\left(\varphi_{t + \tau},\varphi_{t}|H \right)$ is a Gaussian with standard deviation $\beta_H\sigma_H$ and mean value $ \varphi_{t + \tau}^{\text{calc}}\left(\varphi_t,H\right)$, where $\sigma_H$ is the variance of the pointer states, $\beta_H \in (0.8,1.2)$ is a coefficient tuning the filter responsivity, and $ \varphi_{t+\tau}^{\text{calc}}\left(\varphi_t,H\right)$  is the phase corresponding to the classical trajectory starting at a pointer state with $\varphi_{t}$ and ending at the steady state corresponding to $H$ (see Appendix \ref{sec:bayess}).

A typical detection sequence for quantum jumps between the $|\text{e}\rangle,|\text{g}\rangle,$ and $|\text{f}\rangle$ fluxonium states using the recursive Bayesian filter of Eq.\eqref{Bayesian_formula} is shown in  Fig.\ref{bayesian_result}(a).
Based on the measured black trace, the filter estimate for the fluxonium state is indicated by the white trace.
Notice that the $|\text{e}\rangle \rightarrow |\text{f}\rangle$ and $|\text{f}\rangle \rightarrow |\text{e}\rangle$ quantum jumps are correctly identified, even though the pointer state intersects the phase value corresponding to the $|\text{g}\rangle$ steady state.
A quantum jump is declared once one of the probabilities surpasses 50~\%.
The minimum probability for each state is typically capped at 10~\% to prevent the filter saturation at $P_H(t) = 0$, in which case Eq.\eqref{Bayesian_formula} is no longer responsive.
In Fig.\ref{bayesian_result}(b) we selected an example which illustrates the ability of the Bayesian filter to declare a $|\text{g}\rangle \rightarrow|\text{e}\rangle \rightarrow|\text{g}\rangle $ quantum jumps sequence for which the readout resonator never reached the steady state associated with $|\text{e}\rangle$.

In Fig.\ref{fig:detection_time_histos} we compare the Bayesian state estimate to a simpler, more commonly used latching filter (see appendix \ref{section:latching}) for $|\text{g}\rangle \leftrightarrow |\text{e}\rangle$ transition.
In Fig.\ref{fig:detection_time_histos}(a), histograms of the jump detection times are shown for $\tau = 32$~ns.
The Bayesian inference provides a 3.5 times faster state discrimination on average.
In Fig.\ref{fig:detection_time_histos}(b), the mean jump detection time is shown versus $\tau$ for both filters.
As expected, the Bayesian detection starts to outperform the latching filter for $ \tau < \tau_{\text{B}}$ (also see Appendix \ref{sec:bayess}).
The obtained measurement QND fidelity \cite{Touzard2019} $(P_{\text{e}|\text{e}} + P_{\text{g}|\text{g}})/2 = 98 \%$, where $P_{\text{e}|\text{e}}$ and $P_{\text{g}|\text{g}}$ are the probabilities to obtain the same result in  consecutive measurements separated by $\Delta t$, is comparable to state-of-the-art \cite{Jeffrey2014,Krantz2016,Walter2017,Touzard2019,Dassonneville2020}.
The time interval $\Delta t = 432$~ns is chosen such that the detection time of $98\%$ of all quantum jumps is less than $\Delta t$ for the Bayesian filter.
Transitions from $|\text{e}\rangle$ to $|\text{g}\rangle$ and $|\text{f}\rangle$ during $\Delta t$ contribute $1\%$ and $1.2\%$, respectively, to the value of $P_{\text{e}|\text{e}} = 96.7\%$, and transitions from $|\text{g}\rangle$ to $|\text{e}\rangle$ contribute $0.3\%$ to the measured $P_{\text{g}|\text{g}} = 99.6\%$.

In Fig.\ref{fig:detection_time_histos}(c), the average jump detection time for the $|\text{e}\rangle \rightarrow |\text{g}\rangle$ transition is shown as a function of $\overline{n}$.
The results are obtained using the recursive Bayesian filter since the latching filter detecton time saturates at 620~ns (see Fig.\ref{fig:detection_time_histos}(b)).	
The detection time decreases with $\overline{n}$, as expected from Eq.\eqref{eq:SNR}, and it has a minimum of 175~ns for $\overline{n} = 56$.
The increase at $\overline{n} = 110$ is caused mainly by the nonlinearity of the grAl readout resonator and the associated squeezing of the pointer state distributions.

In summary, by exploiting the increasing SNR with $\overline{n}$ we demonstrate a decrease of the artificial atom's state detection time.
This is achieved by monitoring the transient trajectories in the IQ plane of the readout resonator response, triggered by quantum jumps between the three lowest energy states of the artificial atom. 
We show that quantum states can be discriminated by using a recursive Bayesian filter before the resonator reaches its steady state.
In our case the Bayesian filter is applied in post-processing.
However, one can imagine using more sophisticated filtering, and  potentially running it in real time by hardware encoding on a FPGA-based instrument \cite{Gebauer2020, quantum-machines}.
The main limiting factors for the $\overline{n}$ increase are the emergence of non-QND processes, the nonlinearity of the readout resonator, and the saturation of the parametric amplifier.
Thus, mitigating non-QND processes during  readout with increasing photon numbers could further increase the quantum state detection speed in both pulsed and continuous measurements.
\begin{acknowledgments}
Funding was provided by the Alexander von Humboldt foundation in the framework of a Sofja Kovalevskaja award endowed by the German Federal Ministry of Education and Research, and by the Initiative and Networking Fund of the Helmholtz Association, within the Helmholtz Future Project Scalable solid state quantum computing. Furthermore, this research was supported by the ANR under contracts CLOUD (project number ANR-16-CE24-0005). 
IT and AVU acknowledge partial support from the Ministry of Education and Science of the Russian Federation in the framework of the Increase Competitiveness Program of the National University of Science and Technology MISIS (Contract No. K2-2020-017). Facilities use was supported by the KIT Nanostructure Service Laboratory (NSL). 
\end{acknowledgments}
\FloatBarrier
\appendix
\section{Dimer Josephson junction array amplifier (DJJAA)}\label{section:DJJAA}

\begin{figure*}[htbp]
	\includegraphics[width=1. \linewidth] {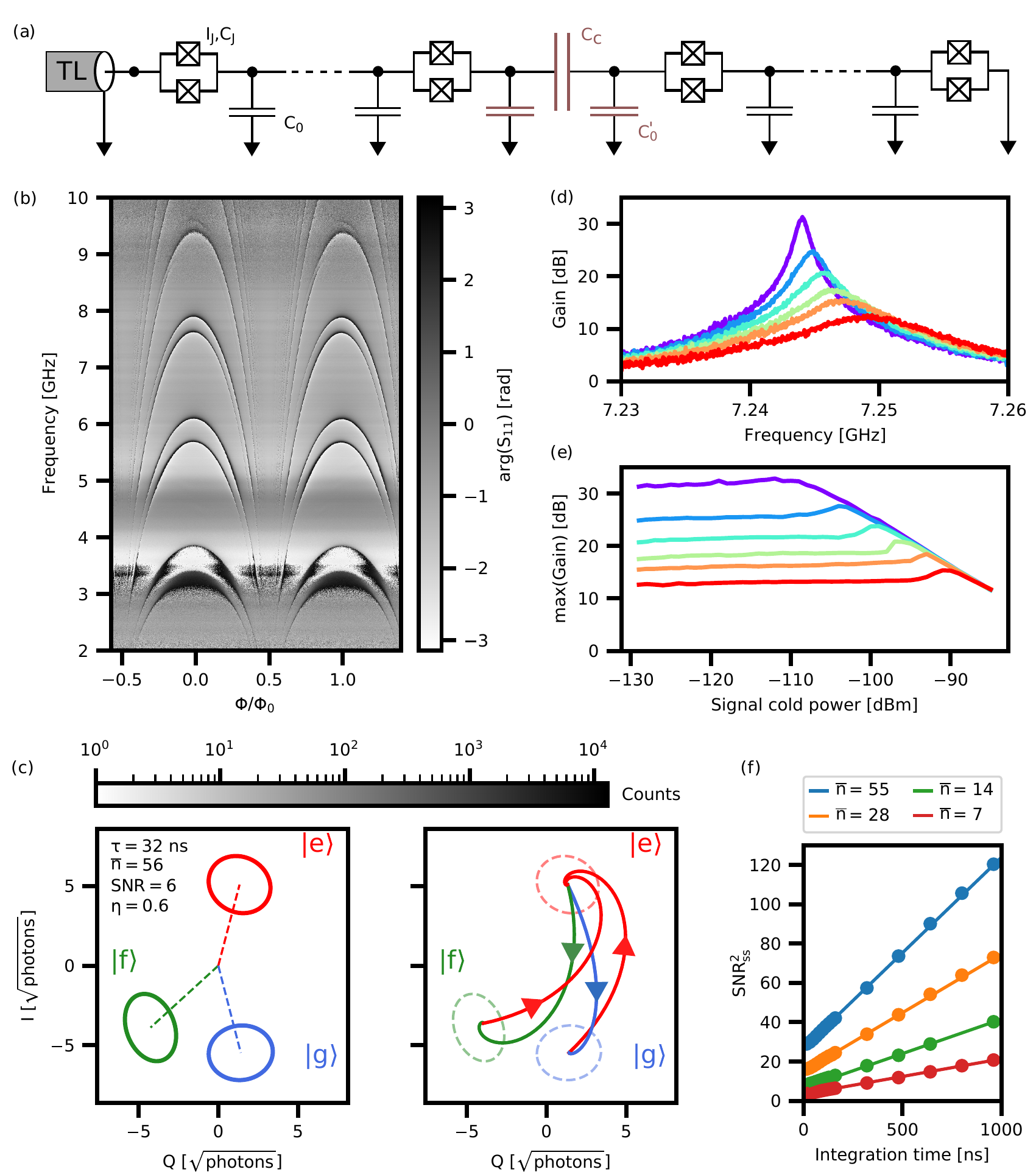}
	\label{fig:DJJAA_image}
	\caption{ 
	(a) Circuit diagram of the DJJAA. 
	(b) Argument of the reflection coefficient of the weak probe signal versus external magnetic flux and frequency.
	(c) 2D histograms of measured I and Q quadratures using the same data as in Fig.\ref{fig:fig1}(c).
	Ellipses indicate the $2\sigma$ areas for $|\text{g}\rangle$, $|\text{e}\rangle$, and $|\text{f}\rangle$ steady states. Dotted lines on the left-hand panel show the location of steady state pointer states $\alpha_i$, with $i \in \{ |\text{g}\rangle, |\text{e}\rangle, |\text{f}\rangle \}$, and arrowed colored curves on the right figure indicate resonator classical trajectories to a steady state of corresponding color. 
	(d) Power gain vs signal frequency for different pump powers. 
	(e) Saturation power of the DJJAA operated at different maximum power gain values.
	(f) Squared steady-state signal-to-noise ratio (Eq.\eqref{eq:SNRss}) as a function of integration time for different $\overline{n}$. 
	Straight lines are linear fits used to extract a measurement efficiency $\eta = 0.6 \pm 0.1$, with the main source of uncertainty being the photon number calibration.
	}
\end{figure*}

We utilize a dimer Josephson junction array amplifier (DJJAA) \cite{Winkel2019} which consists of an array of $N = 1600$ SQUIDs interrupted in the middle by an interdigitated capacitor (see Fig.\ref{fig:DJJAA_image}(a)).
The capacitor breaks the symmetry between odd and even modes and creates pairs of modes (see Fig.\ref{fig:DJJAA_image}(b)).
Each of these pairs, which we refer to as dimers, can be used for 4 wave-mixing non-degenerate parametric amplification.
In this regime signal and idler tones occupy different physical modes, and the gain profile has a double-Lorentzian shape when the pump tone is applied in-between the modes.

The parameters of the DJJAA (Sample N=1600 reported in Ref.\cite{Winkel2019}) are: $I_{\text{J}}^{\text{SQUID}} = 3.5$ \textmu A, $C_{\text{J}} = 1225$~fF, $C_0 = 0.45$~fF, $C_{\text{c}} = 40$~fF, and $C_0' = 33$~fF.
The power gains presented in Fig.\ref{fig:DJJAA_image}(d) are the lower frequency lobes of the non-degenerate gain profiles (the pump power dependent lobes splitting is $260\pm 20$~MHz).
For 20 dB of gain we obtained an instantaneous band of 7 MHz, and a saturation power of -98~dBm (see Fig.\ref{fig:DJJAA_image}(e)).

The measurement efficiency $\eta$ was extracted by fitting the dependence of the steady-state signal-to-noise ratio with the integration time
\begin{equation}
\text{SNR}_{\text{ss}}= \frac{|\bf{\alpha}|}{\sqrt{\sigma_Q^2 + \sigma_I^2}} = \sqrt{\kappa B^{-1} \eta \overline{n} /4},
\label{eq:SNRss}
\end{equation}
where $\bf{\alpha}$ is the pointer state of the ground state and $\sigma_{I/Q}$ are it's variances along respective quadratures, see Fig.\ref{fig:DJJAA_image}(c)).
$B^{-1}$ is the measurement bandwidth (see supplementary material in Ref.\cite{Vijay2011}), which depends on all the measurement setup bandwidths, including integration time $\tau$, readout resonator linewidth $\kappa$, DJJAA bandwidth, etc.
In our case the main limiting factors are $\tau$ and $\kappa$, therefore we can consider $B^{-1} \approx \tau + \tau_B$, where $\tau_B = 2/\kappa$.
The $2/\kappa$ comes from the readout resonator pointer state evolution $\alpha(t) - \alpha_{\text{f}} \propto \left( \alpha_{\text{i}} - \alpha_{\text{f}}\right)e^{-\kappa t /2}$, where $\alpha_{\text{f/i}}$ are the mean pointer states of final and initial steady states (see Appendix \ref{sec:bayess}).

\begin{figure}[!t]
	\includegraphics[width=1 \linewidth] {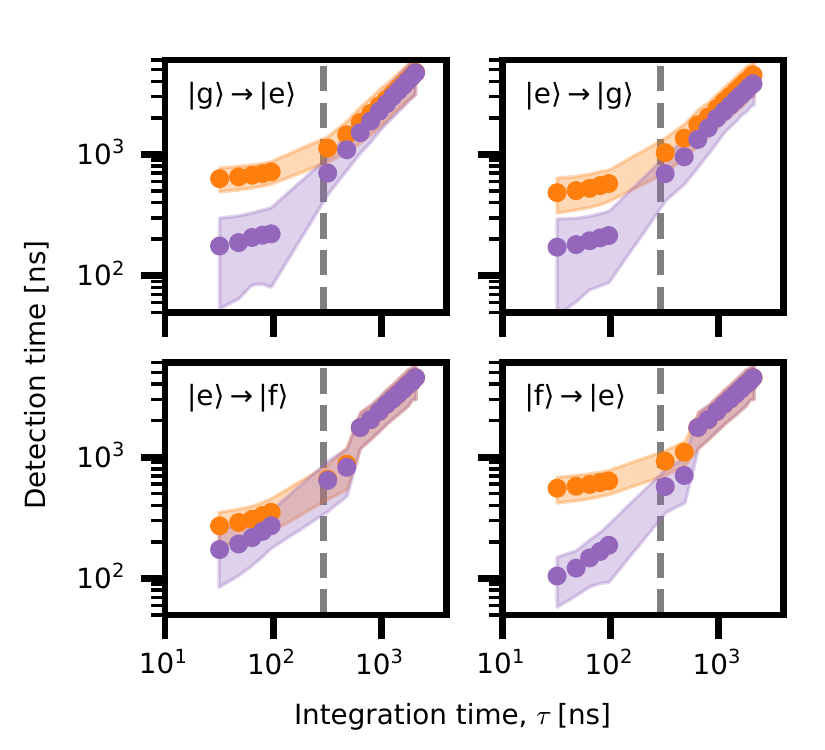}
	\label{fig:all_detection_times}
	\caption{ 
		Average detection time for $|\text{g}\rangle \rightarrow |\text{e}\rangle$,$|\text{e}\rangle \rightarrow |\text{g}\rangle$,$|\text{e}\rangle \rightarrow |\text{f}\rangle$,$|\text{f}\rangle \rightarrow |\text{e}\rangle$ jumps as a function of integration time, obtained using the three-point latching (in orange) or the recursive Bayesian filter (in purple).
		Transparent areas represent the standard deviation of the jump detection time distribution.
		The grey dashed line corresponds to $\tau_{\text{B}}$.}
\end{figure}
\section{Latching filtering} \label{section:latching}
The latching filter was designed to declare a jump when a pointer state enters the $2\sigma$ area (Fig.\ref{fig:DJJAA_image}(c)) of a respective steady state.
The detection time (for both latching and bayesian filtering) is defined as time between leaving the $2\sigma$ area of the previous coherent state and jump detection. 

The steady-states are visibly squeezed because the readout signal effectively acts as a pump of a parametric amplifier, squeezing the vacuum noise.
This happens thanks to the grAl readout resonator's intrinsic nonlinearity \cite{Maleeva2018} $K^{\text{grAl}}/2\pi = - 2.4$~kHz and the inherited state-dependent nonlinearities \cite{Smith2016} $K^{|\text{g}\rangle}/2\pi = -2.6$~Hz, $K^{|\text{e}\rangle}/2\pi = 2$~kHz, and $K^{|\text{f}\rangle}/2\pi = -2$~kHz at $\overline{n} = 50$ (see Ref.\cite{Gusenkova2020}).
Notice that for the $|\text{e}\rangle$ state the Kerr coefficients almost cancel out, as evidenced by the reduced squeezing of the $|\text{e}\rangle$ pointer state in Fig.\ref{fig:DJJAA_image}(c). 
\section{Bayesian filtering}\label{sec:bayess}
The state of the quantum system is encoded in the instantaneous evolution of the pointer state.
The pointer state follows classical trajectories, and the expected position of the pointer state $\alpha_{t + \tau}^{\text{calc}}(\alpha_t, H)$ can be calculated for each fluxonium state hypothesis $H$:
\begin{multline}
\alpha_{t + \tau}^{\text{calc}}(\alpha_t, H) = \\
(\alpha_t - A_H)e^{-\kappa \tau / 2 + i(\omega_{\text{drive}} - \omega_H )\tau  } + A_H
\label{eq:complex_plane_eq}
\end{multline}
Here $\alpha_t$ is the measured pointer state at time t, $A_H$ is the steady state of the system, $\omega_{\text{drive}}$ is the readout drive frequency, and $\omega_H$ is the state-dependent readout resonator frequency.
This expression can be used for the calculation of the pointer state evolution for the Bayesian inference (Eq.\eqref{Bayesian_formula_explanation}).
However, if the angle between pointer states is big enough, it is also possible to make a more robust approximation and only track the phase (or a quadrature), assuming that it evolves as 
\begin{equation}
\varphi_{t + \tau}^{\text{calc}}(\varphi_t, H) \approx (\varphi_t - \phi_H)e^{-C_H\kappa \tau / 2 } + \phi_H
\label{eq:phase_eq}
\end{equation}
Here $\varphi_t$ is the measured phase at time $t$, $\phi_H$ is the phase of the respective steady state, and $C_H = 1 \pm 0.2$ an empirical coefficient accounting for the resonator nonlinearity and for the term $e^{i(\omega_{\text{drive}} - \omega_H )\tau}$ in Eq.\eqref{eq:complex_plane_eq}.
Both pointer state evolution (Eq.\eqref{eq:complex_plane_eq}) and exponential phase evolution (Eq.\eqref{eq:phase_eq}) showed similar results when used for the Bayesian filtering, and in this work  we utilized the simpler Eq.\eqref{eq:phase_eq} to calculate the expected phase $\varphi_{t +\tau}^{\text{calc}}$ for Eq.\eqref{Bayesian_formula_explanation}.

\vfill\null
\section{Measurement setup}\label{sec:measurement_setup}
\begin{figure}[!t]
	\includegraphics[width=1 \linewidth] {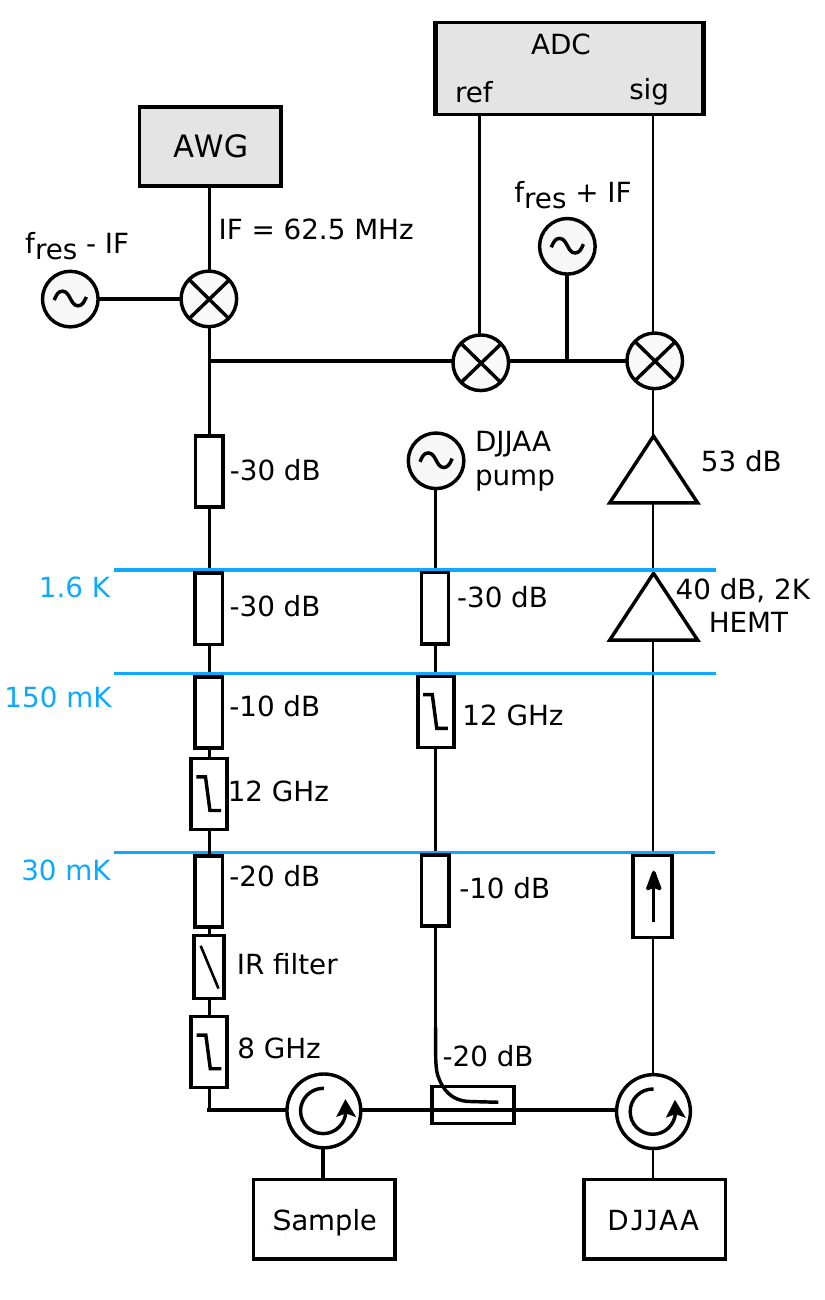}
	\label{fig:setup_scheme}
	\caption{ 
	Schematics of the measurement setup.
	The displayed microwave components are thermalized to the nearest temperature stage indicated above them.
	}
\end{figure}

The measurement setup is shown in Fig.\ref{fig:setup_scheme}.
In front of the ADC we use passband filters with a bandwidth of 10~MHz.
An additional post-processing filtering with a bandwidth of 4~MHz was applied to the signal trace.
\FloatBarrier
\bibliography{bibliography}

\begin{thebibliography}{10}

\bibitem{Nagourney1986}
W.~Nagourney, J.~Sandberg, and H.~Dehmelt, ``Shelved optical electron
  amplifier: Observation of quantum jumps,'' {\em Phys. Rev. Lett.}, vol.~56,
  pp.~2797--2799, Jun 1986.

\bibitem{Neumann2010}
P.~Neumann, J.~Beck, M.~Steiner, F.~Rempp, H.~Fedder, P.~R. Hemmer,
  J.~Wrachtrup, and F.~Jelezko, ``Single-shot readout of a single nuclear
  spin,'' {\em Science}, vol.~329, no.~5991, pp.~542--544, 2010.

\bibitem{Robledo2011}
L.~Robledo, L.~Childress, H.~Bernien, B.~Hensen, P.~F.~A. Alkemade, and
  R.~Hanson, ``High-fidelity projective read-out of a solid-state spin quantum
  register,'' {\em Nature}, vol.~477, pp.~574--578, September 2011.

\bibitem{Gleyzes2007}
S.~Gleyzes, S.~Kuhr, C.~Guerlin, J.~Bernu, S.~Deléglise, U.~B. Hoff, M.~Brune,
  J.-M. Raimond, and S.~Haroche, ``Quantum jumps of light recording the birth
  and death of a photon in a cavity,'' {\em Nature}, vol.~446, p.~297—300,
  March 2007.

\bibitem{Jelezko2002}
F.~Jelezko, I.~Popa, A.~Gruber, C.~Tietz, J.~Wrachtrup, A.~Nizovtsev, and
  S.~Kilin, ``Single spin states in a defect center resolved by optical
  spectroscopy,'' {\em Applied Physics Letters}, vol.~81, no.~12,
  pp.~2160--2162, 2002.

\bibitem{Vijay2011}
R.~Vijay, D.~H. Slichter, and I.~Siddiqi, ``Observation of quantum jumps in a
  superconducting artificial atom,'' {\em Phys. Rev. Lett.}, vol.~106,
  p.~110502, Mar 2011.

\bibitem{Reed2012}
M.~D. Reed, L.~DiCarlo, S.~E. Nigg, L.~Sun, L.~Frunzio, S.~M. Girvin, and R.~J.
  Schoelkopf, ``Realization of three-qubit quantum error correction with
  superconducting circuits,'' {\em Nature}, vol.~482, p.~382–385, Feb 2012.

\bibitem{Riste2013a}
D.~Ristè, M.~Dukalski, C.~A. Watson, G.~de~Lange, M.~J. Tiggelman, Y.~M.
  Blanter, K.~W. Lehnert, R.~N. Schouten, and L.~DiCarlo, ``Deterministic
  entanglement of superconducting qubits by parity measurement and feedback,''
  {\em Nature}, vol.~502, pp.~350--354, Feb 2012.

\bibitem{Sun2014}
L.~Sun, A.~Petrenko, Z.~Leghtas, B.~Vlastakis, G.~Kirchmair, K.~M. Sliwa,
  A.~Narla, M.~Hatridge, S.~Shankar, J.~Blumoff, L.~Frunzio, M.~Mirrahimi,
  M.~H. Devoret, and R.~J. Schoelkopf, ``Tracking photon jumps with repeated
  quantum non-demolition parity measurements,'' {\em Nature}, vol.~511,
  p.~444–448, Jul 2014.

\bibitem{Kelly2015}
J.~Kelly, R.~Barends, A.~G. Fowler, A.~Megrant, E.~Jeffrey, T.~C. White,
  D.~Sank, J.~Y. Mutus, B.~Campbell, Y.~Chen, Z.~Chen, B.~Chiaro, A.~Dunsworth,
  I.~C. Hoi, C.~Neill, P.~J.~J. O'Malley, C.~Quintana, P.~Roushan,
  A.~Vainsencher, J.~Wenner, A.~N. Cleland, and J.~M. Martinis, ``State
  preservation by repetitive error detection in a superconducting quantum
  circuit,'' {\em Nature}, vol.~519, p.~66–69, Mar 2015.

\bibitem{Ofek2016}
N.~Ofek, A.~Petrenko, R.~Heeres, P.~Reinhold, Z.~Leghtas, B.~Vlastakis, Y.~Liu,
  L.~Frunzio, S.~Girvin, L.~Jiang, M.~Mirrahimi, M.~Devoret, and R.~Schoelkopf,
  ``Extending the lifetime of a quantum bit with error correction in
  superconducting circuits,'' {\em Nature}, vol.~536, p.~441—445, August
  2016.

\bibitem{Riste2013}
D.~Riste, C.~C. Bultink, M.~J. Tiggelman, R.~N. Schouten, K.~W. Lehnert, and
  L.~DiCarlo, ``Millisecond charge-parity fluctuations and induced decoherence
  in a superconducting transmon qubit,'' {\em Nature Communications}, vol.~4,
  May 2013.

\bibitem{Vool2014}
U.~Vool, I.~M. Pop, K.~Sliwa, B.~Abdo, C.~Wang, T.~Brecht, Y.~Y. Gao,
  S.~Shankar, M.~Hatridge, G.~Catelani, M.~Mirrahimi, L.~Frunzio, R.~J.
  Schoelkopf, L.~I. Glazman, and M.~H. Devoret, ``Non-poissonian quantum jumps
  of a fluxonium qubit due to quasiparticle excitations,'' {\em Phys. Rev.
  Lett.}, vol.~113, p.~247001, Dec 2014.

\bibitem{Serniak2019}
K.~Serniak, S.~Diamond, M.~Hays, V.~Fatemi, S.~Shankar, L.~Frunzio,
  R.~Schoelkopf, and M.~Devoret, ``Direct dispersive monitoring of charge
  parity in offset-charge-sensitive transmons,'' {\em Physical Review Applied},
  vol.~12, Jul 2019.

\bibitem{Wallraff2004}
A.~Wallraff, D.~I. Schuster, A.~Blais, L.~Frunzio, R.~S. Huang, J.~Majer,
  S.~Kumar, S.~M. Girvin, and R.~J. Schoelkopf, ``Strong coupling of a single
  photon to a superconducting qubit using circuit quantum electrodynamics,''
  {\em Nature}, vol.~431, p.~162–167, Sep 2004.

\bibitem{Krantz2019a}
P.~Krantz, M.~Kjaergaard, F.~Yan, T.~P. Orlando, S.~Gustavsson, and W.~D.
  Oliver, ``{A quantum engineer's guide to superconducting qubits},'' {\em
  Applied Physics Reviews}, vol.~6, no.~2, 2019.

\bibitem{Blais2020a}
A.~Blais, S.~M. Girvin, and W.~D. Oliver, ``{Quantum information processing and
  quantum optics with circuit quantum electrodynamics},'' {\em Nature Physics},
  vol.~16, no.~3, pp.~247--256, 2020.

\bibitem{Kjaergaard2020}
M.~Kjaergaard, M.~E. Schwartz, J.~Braum{\"{u}}ller, P.~Krantz, J.~I. Wang,
  S.~Gustavsson, and W.~D. Oliver, ``{Superconducting Qubits: Current State of
  Play},'' {\em Annual Review of Condensed Matter Physics}, vol.~11,
  pp.~369--395, 2020.

\bibitem{Jeffrey2014}
E.~Jeffrey, D.~Sank, J.~Y. Mutus, T.~C. White, J.~Kelly, R.~Barends, Y.~Chen,
  Z.~Chen, B.~Chiaro, A.~Dunsworth, A.~Megrant, P.~J.~J. O'Malley, C.~Neill,
  P.~Roushan, A.~Vainsencher, J.~Wenner, A.~N. Cleland, and J.~M. Martinis,
  ``Fast accurate state measurement with superconducting qubits,'' {\em Phys.
  Rev. Lett.}, vol.~112, p.~190504, May 2014.

\bibitem{Walter2017}
T.~Walter, P.~Kurpiers, S.~Gasparinetti, P.~Magnard,
  A.~Poto\ifmmode~\check{c}\else \v{c}\fi{}nik, Y.~Salath\'e, M.~Pechal,
  M.~Mondal, M.~Oppliger, C.~Eichler, and A.~Wallraff, ``Rapid high-fidelity
  single-shot dispersive readout of superconducting qubits,'' {\em Phys. Rev.
  Applied}, vol.~7, p.~054020, May 2017.

\bibitem{Dassonneville2020}
R.~Dassonneville, T.~Ramos, V.~Milchakov, L.~Planat, .~Dumur, F.~Foroughi,
  J.~Puertas, S.~Leger, K.~Bharadwaj, J.~Delaforce, C.~Naud, W.~Hasch-Guichard,
  J.~J. García-Ripoll, N.~Roch, and O.~Buisson, ``Fast high-fidelity quantum
  nondemolition qubit readout via a nonperturbative cross-kerr coupling,'' {\em
  Physical Review X}, vol.~10, Feb 2020.

\bibitem{Castellanos-Beltran2007}
M.~A. Castellanos-Beltran and K.~W. Lehnert, ``Widely tunable parametric
  amplifier based on a superconducting quantum interference device array
  resonator,'' {\em Applied Physics Letters}, vol.~91, p.~083509, Aug 2007.

\bibitem{Yamamoto2008}
T.~Yamamoto, K.~Inomata, M.~Watanabe, K.~Matsuba, T.~Miyazaki, W.~D. Oliver,
  Y.~Nakamura, and J.~S. Tsai, ``Flux-driven josephson parametric amplifier,''
  {\em Applied Physics Letters}, vol.~93, no.~4, p.~042510, 2008.

\bibitem{Mutus2014}
J.~Mutus, T.~White, R.~Barends, Y.~Chen, Z.~Chen, B.~Chiaro, A.~Dunsworth,
  E.~Jeffrey, J.~Kelly, A.~Megrant, C.~Neill, P.~O'Malley, P.~Roushan, D.~Sank,
  A.~Vainsencher, J.~Wenner, K.~Sundqvist, A.~Cleland, and J.~Martinis,
  ``Strong environmental coupling in a josephson parametric amplifier,'' {\em
  Applied Physics Letters}, vol.~104, p.~263513, Jun 2014.

\bibitem{Roch2012}
N.~Roch, E.~Flurin, F.~Nguyen, P.~Morfin, P.~Campagne-Ibarcq, M.~H. Devoret,
  and B.~Huard, ``Widely tunable, nondegenerate three-wave mixing microwave
  device operating near the quantum limit,'' {\em Physical Review Letters},
  vol.~108, Apr 2012.

\bibitem{Eichler2014}
C.~Eichler and A.~Wallraff, ``Controlling the dynamic range of a josephson
  parametric amplifier,'' {\em EPJ Quantum Technology}, vol.~1, Jan 2014.

\bibitem{Macklin2015}
C.~Macklin, K.~O'Brien, D.~Hover, M.~Schwartz, V.~Bolkhovsky, X.~Zhang,
  W.~Oliver, and I.~Siddiqi, ``A near-quantum-limited josephson traveling-wave
  parametric amplifier,'' {\em Science (New York, N.Y.)}, vol.~350,
  p.~307—310, October 2015.

\bibitem{Roy2016}
A.~Roy and M.~Devoret, ``Introduction to parametric amplification of quantum
  signals with josephson circuits,'' {\em Comptes Rendus Physique}, vol.~17,
  p.~740–755, Aug 2016.

\bibitem{Caves1982}
C.~M. Caves, ``Quantum limits on noise in linear amplifiers,'' {\em Phys. Rev.
  D}, vol.~26, pp.~1817--1839, Oct 1982.

\bibitem{Johnson2012}
J.~E. Johnson, C.~Macklin, D.~H. Slichter, R.~Vijay, E.~B. Weingarten,
  J.~Clarke, and I.~Siddiqi, ``Heralded state preparation in a superconducting
  qubit,'' {\em Physical Review Letters}, vol.~109, Aug 2012.

\bibitem{Sank2016}
D.~Sank, Z.~Chen, M.~Khezri, J.~Kelly, R.~Barends, B.~Campbell, Y.~Chen,
  B.~Chiaro, A.~Dunsworth, A.~Fowler, E.~Jeffrey, E.~Lucero, A.~Megrant,
  J.~Mutus, M.~Neeley, C.~Neill, P.~J.~J. O'Malley, C.~Quintana, P.~Roushan,
  A.~Vainsencher, T.~White, J.~Wenner, A.~N. Korotkov, and J.~M. Martinis,
  ``Measurement-induced state transitions in a superconducting qubit: Beyond
  the rotating wave approximation,'' {\em Phys. Rev. Lett.}, vol.~117,
  p.~190503, Nov 2016.

\bibitem{Minev2019}
Z.~K. Minev, S.~O. Mundhada, S.~Shankar, P.~Reinhold, R.~Gutiérrez-Jáuregui,
  R.~J. Schoelkopf, M.~Mirrahimi, H.~J. Carmichael, and M.~H. Devoret, ``To
  catch and reverse a quantum jump mid-flight,'' {\em Nature}, vol.~570,
  p.~200–204, Jun 2019.

\bibitem{Boissonneault2009}
M.~Boissonneault, J.~M. Gambetta, and A.~Blais, ``Dispersive regime of circuit
  {QED}: Photon-dependent qubit dephasing and relaxation rates,'' {\em Physical
  Review A}, vol.~79, Jan 2009.

\bibitem{Malekakhlagh2020}
M.~Malekakhlagh, A.~Petrescu, and H.~E. Türeci, ``Lifetime renormalization of
  weakly anharmonic superconducting qubits. i. role of number nonconserving
  terms,'' {\em Physical Review B}, vol.~101, Apr 2020.

\bibitem{Petrescu2020}
A.~Petrescu, M.~Malekakhlagh, and H.~E. Türeci, ``Lifetime renormalization of
  driven weakly anharmonic superconducting qubits. ii. the readout problem,''
  {\em Physical Review B}, vol.~101, Apr 2020.

\bibitem{Grunhaupt2019}
L.~Grünhaupt, M.~Spiecker, D.~Gusenkova, N.~Maleeva, S.~T. Skacel,
  I.~Takmakov, F.~Valenti, P.~Winkel, H.~Rotzinger, A.~V. Ustinov, and I.~M.
  Pop, ``Granular aluminium as a superconducting material for high-impedance
  quantum circuits,'' {\em Nature Materials}, vol.~18, p.~816–819, Apr 2019.

\bibitem{Gusenkova2020}
D.~Gusenkova, M.~Spiecker, R.~Gebauer, M.~Willsch, F.~Valenti, N.~Karcher,
  L.~Grünhaupt, I.~Takmakov, P.~Winkel, D.~Rieger, A.~V. Ustinov, N.~Roch,
  W.~Wernsdorfer, K.~Michielsen, O.~Sander, and I.~M. Pop, ``Quantum
  non-demolition dispersive readout of a superconducting artificial atom using
  large photon numbers,'' 2020.

\bibitem{Winkel2019}
P.~Winkel, I.~Takmakov, D.~Rieger, L.~Planat, W.~Hasch-Guichard,
  L.~Gr\"unhaupt, N.~Maleeva, F.~Foroughi, F.~Henriques, K.~Borisov,
  J.~Ferrero, A.~V. Ustinov, W.~Wernsdorfer, N.~Roch, and I.~M. Pop,
  ``Nondegenerate parametric amplifiers based on dispersion-engineered
  josephson-junction arrays,'' {\em Phys. Rev. Applied}, vol.~13, p.~024015,
  Feb 2020.

\bibitem{Korotkov2016}
A.~N. Korotkov, ``Quantum bayesian approach to circuit {QED} measurement with
  moderate bandwidth,'' {\em Physical Review A}, vol.~94, Oct 2016.

\bibitem{Wang2015}
P.~Wang, L.~Qin, and X.-Q. Li, ``Quantum bayesian rule for weak measurements of
  qubits in superconducting circuit {QED},'' {\em New Journal of Physics},
  vol.~16, p.~123047, Dec 2014.

\bibitem{Weber2016a}
S.~J. Weber, K.~W. Murch, M.~E. Kimchi-Schwartz, N.~Roch, and I.~Siddiqi,
  ``Quantum trajectories of superconducting qubits,'' {\em Comptes Rendus
  Physique}, vol.~17, p.~766–777, Aug 2016.

\bibitem{Feng2016}
W.~Feng, P.~Liang, L.~Qin, and X.-Q. Li, ``Exact quantum bayesian rule for
  qubit measurements in circuit {QED},'' {\em Scientific Reports}, vol.~6, Feb
  2016.

\bibitem{Reick2010}
S.~Reick, K.~Mølmer, W.~Alt, M.~Eckstein, T.~Kampschulte, L.~Kong, R.~Reimann,
  A.~Thobe, A.~Widera, and D.~Meschede, ``Analyzing quantum jumps of one and
  two atoms strongly coupled to an optical cavity,'' {\em Journal of the
  Optical Society of America B}, vol.~27, p.~A152, May 2010.

\bibitem{Six2016}
P.~Six, P.~Campagne-Ibarcq, I.~Dotsenko, A.~Sarlette, B.~Huard, and P.~Rouchon,
  ``Quantum state tomography with noninstantaneous measurements, imperfections,
  and decoherence,'' {\em Physical Review A}, vol.~93, Jan 2016.

\bibitem{Manucharyan2009}
V.~E. Manucharyan, J.~Koch, L.~I. Glazman, and M.~H. Devoret, ``Fluxonium:
  Single cooper-pair circuit free of charge offsets,'' {\em Science}, vol.~326,
  p.~113–116, Oct 2009.

\bibitem{Lin2018}
Y.-H. Lin, L.~B. Nguyen, N.~Grabon, J.~San~Miguel, N.~Pankratova, and V.~E.
  Manucharyan, ``Demonstration of protection of a superconducting qubit from
  energy decay,'' {\em Physical Review Letters}, vol.~120, Apr 2018.

\bibitem{Kou2018}
A.~Kou, W.~Smith, U.~Vool, I.~Pop, K.~Sliwa, M.~Hatridge, L.~Frunzio, and
  M.~Devoret, ``Simultaneous monitoring of fluxonium qubits in a waveguide,''
  {\em Physical Review Applied}, vol.~9, Jun 2018.

\bibitem{Eichler2014a}
C.~Eichler, Y.~Salathe, J.~Mlynek, S.~Schmidt, and A.~Wallraff,
  ``Quantum-limited amplification and entanglement in coupled nonlinear
  resonators,'' {\em Physical Review Letters}, vol.~113, Sep 2014.

\bibitem{Maleeva2018}
N.~Maleeva, L.~Grünhaupt, T.~Klein, F.~Levy-Bertrand, O.~Dupré, M.~Calvo,
  F.~Valenti, P.~Winkel, F.~Friedrich, W.~Wernsdorfer, A.~V. Ustinov,
  H.~Rotzinger, A.~Monfardini, M.~V. Fistul, and I.~M. Pop, ``Circuit quantum
  electrodynamics of granular aluminum resonators,'' {\em Nature
  Communications}, vol.~9, Sep 2018.

\bibitem{Touzard2019}
S.~Touzard, A.~Kou, N.~Frattini, V.~Sivak, S.~Puri, A.~Grimm, L.~Frunzio,
  S.~Shankar, and M.~Devoret, ``Gated conditional displacement readout of
  superconducting qubits,'' {\em Physical Review Letters}, vol.~122, Feb 2019.

\bibitem{Krantz2016}
P.~Krantz, A.~Bengtsson, M.~Simoen, S.~Gustavsson, V.~Shumeiko, W.~D. Oliver,
  C.~M. Wilson, P.~Delsing, and J.~Bylander, ``Single-shot read-out of a
  superconducting qubit using a josephson parametric oscillator,'' {\em Nature
  Communications}, vol.~7, May 2016.

\bibitem{Gebauer2020}
R.~Gebauer, N.~Karcher, D.~Gusenkova, M.~Spiecker, L.~Grünhaupt, I.~Takmakov,
  P.~Winkel, L.~Planat, N.~Roch, W.~Wernsdorfer, A.~V. Ustinov, M.~Weber,
  M.~Weides, I.~M. Pop, and O.~Sander, ``State preparation of a fluxonium qubit
  with feedback from a custom fpga-based platform,'' {\em AIP Conference
  Proceedings}, vol.~2241, no.~1, p.~020015, 2020.

\bibitem{quantum-machines}
\url{https://www.quantum-machines.com}.

\bibitem{Smith2016}
W.~C. Smith, A.~Kou, U.~Vool, I.~M. Pop, L.~Frunzio, R.~J. Schoelkopf, and
  M.~H. Devoret, ``Quantization of inductively shunted superconducting
  circuits,'' {\em Physical Review B}, vol.~94, Oct 2016.

\end{thebibliography}
\bibliographystyle{ieeetr}
\end{document}